\documentclass[]{aastex631}
\usepackage{xspace}

\newcommand{\logg}{\ensuremath{\log{g}}\xspace}
\newcommand{\vsini}{\ensuremath{v \sin i}\xspace} 

\newcommand{\Msun}{\ensuremath{M_{\odot}}\xspace}

\newcommand{\Teff}{\ensuremath{T_\mathrm{eff}}\xspace}

\newcommand{\ed}{\textcolor{black}}

\newcommand{\caltech}{Department of Astronomy, California Institute of Technology, Pasadena, CA 91125, USA}
\newcommand{\northwestern}{Center for Interdisciplinary Exploration and Research in Astrophysics (CIERA) and Department of Physics and Astronomy, Northwestern University, Evanston, IL 60208, USA}

\newcommand{\ucsc}{Department of Astronomy \& Astrophysics, University of California, Santa Cruz, CA95064, USA}
\newcommand{\keck}{W. M. Keck Observatory, 65-1120 Mamalahoa Hwy, Kamuela, HI, USA}
\newcommand{\ucla}{Department of Physics \& Astronomy, 430 Portola Plaza, University of California, Los Angeles, CA 90095, USA}
\newcommand{\jpl}{Jet Propulsion Laboratory, California Institute of Technology, 4800 Oak Grove Dr.,Pasadena, CA 91109, USA}
\newcommand{\ucsd}{Department of Astronomy and Astrophysics, University of California, San Diego, La Jolla, CA 92093}

\newcommand{\ucsb}{Department of Physics, University of California, Santa Barbara, CA 93106, USA}
\newcommand{\osu}{Department of Astronomy, The Ohio State University, Columbus, OH 43210, USA}
\newcommand{\UofA}{James C. Wyant College of Optical Sciences, University of Arizona, Meinel Building 1630 E. University Blvd., Tucson, AZ. 85721}

\newcommand{\UoE}{School of Physics and Astronomy, University of Exeter, Exeter, EX4 4QL, UK}
\newcommand{\chara}{The CHARA Array of Georgia State University, Mount Wilson Observatory, Mount Wilson, CA 91023, USA}

\submitjournal{ApJ Letters}

\usepackage{multirow}       
\usepackage{tablefootnote}  
\usepackage{makecell}       
\usepackage{enumitem}       

\begin{document}

\title{Vortex Fiber Nulling for Exoplanet Observations: First Direct Detection of M Dwarf Companions around HIP~21543, HIP~94666, and HIP~50319}


\correspondingauthor{Daniel Echeverri}
\email{dechever@caltech.edu}

\author[0000-0002-1583-2040]{Daniel Echeverri}
\affiliation{\caltech}

\author[0000-0002-6618-1137]{Jerry W. Xuan}
\affiliation{\caltech}

\author[0000-0002-3380-3307]{John D. Monnier}
\affiliation{Astronomy Department, University of Michigan, Ann Arbor, MI 48109, USA}

\author[0000-0001-8953-1008]{Jacques-Robert Delorme}
\affiliation{\keck}

\author[0000-0003-0774-6502]{Jason J. Wang}
\affiliation{\northwestern}

\author[0000-0001-5213-6207]{Nemanja Jovanovic}
\affiliation{\caltech}

\author[0000-0001-9708-8667]{Katelyn Horstman}
\affiliation{\caltech}

\author[0000-0003-4769-1665]{Garreth Ruane}
\affiliation{\jpl}

\author{Bertrand Mennesson}
\affiliation{\jpl}

\author{Eugene Serabyn}
\affiliation{\jpl}

\author[0000-0002-8895-4735]{Dimitri Mawet}
\affiliation{\caltech}
\affiliation{\jpl}

\author{J. Kent Wallace}
\affiliation{\jpl}

\author{Sofia Hillman}
\affiliation{\ucsb}

\author{Ashley Baker}
\affiliation{\caltech}

\author{Randall Bartos}
\affiliation{\jpl}

\author{Benjamin Calvin}
\affiliation{\ucla}

\author{Sylvain Cetre}
\affiliation{\keck}

\author{Greg Doppmann}
\affiliation{\keck}

\author[0000-0002-1392-0768]{Luke Finnerty}
\affiliation{\ucla}

\author[0000-0002-0176-8973]{Michael P. Fitzgerald}
\affiliation{\ucla}

\author[0000-0002-5370-7494]{Chih-Chun Hsu}
\affiliation{\northwestern}

\author[0000-0002-4934-3042]{Joshua Liberman}
\affiliation{\UofA}

\author{Ronald L\'opez}
\affiliation{\ucla}

\author{Maxwell Millar-Blanchaer}
\affiliation{\ucsb}

\author{Evan Morris}
\affiliation{\ucsc}

\author{Jacklyn Pezzato}
\affiliation{\caltech}

\author[0000-0003-2233-4821]{Jean-Baptiste Ruffio}
\affiliation{\ucsd}

\author{Ben Sappey}
\affiliation{\ucsd}

\author{Tobias Schofield}
\affiliation{\caltech}

\author{Andrew J. Skemer}
\affiliation{\ucsc}

\author{Ji Wang}
\affiliation{\osu}

\author{Yinzi Xin}
\affiliation{\caltech}


\author[0000-0002-2208-6541]{Narsireddy Anugu}
\affiliation{\chara}

\author[0000-0001-8926-9732]{Sorabh Chhabra}
\affiliation{\UoE}

\author[0000-0002-1788-9366]{Noura Ibrahim}
\affiliation{Astronomy Department, University of Michigan, Ann Arbor, MI 48109, USA}

\author{Stefan Kraus}
\affiliation{\UoE}

\author[0000-0001-5415-9189]{Gail H. Schaefer}
\affiliation{\chara}

\author[0000-0001-9745-5834]{Cyprien Lanthermann}
\affiliation{\chara}


\begin{abstract}
Vortex fiber nulling (VFN) is a technique for detecting and characterizing faint companions at small separations from their host star. A near-infrared (${\sim}2.3~\mu$m) VFN demonstrator mode was deployed on the Keck Planet Imager and Characterizer (KPIC) instrument at the Keck Observatory and presented earlier. In this paper, we present the first VFN companion detections. Three targets, HIP~21543~Ab, HIP~94666~Ab, and HIP~50319~B, were detected with host-companion flux ratios between 70 and 430 at and within one diffraction beamwidth \ed{($\lambda/D$)}. 
We complement the spectra from KPIC VFN with flux ratio and position measurements from the CHARA Array to validate the VFN results and provide a more complete characterization of the targets. 
This paper reports the first direct detection of these three M dwarf companions, yielding their first spectra and flux ratios. Our observations provide measurements of bulk properties such as effective temperatures, radial velocities, and v$\sin{i}$, and verify the accuracy of the published orbits. These detections corroborate earlier predictions of the KPIC VFN performance, demonstrating that the instrument mode is ready for science observations.
\end{abstract}


\section{Introduction} \label{sec:intro}



Decades of radial velocity (RV) surveys have revealed that giant planets are most likely to orbit between 1-10~AU from their host stars \citep{Fulton2021_OccurRates,Rosenthal2021_OccurRates}. \ed{However, typical coronagraphs have inner working angles (IWA) of about $3~\lambda/D$ \citep{Macintosh2014_GPIInstrument,Beuzit2019_SPHEREInstrument}, so they cannot efficiently observe exoplanets at small separations. Here $\lambda$ is the operating wavelength and $D$ is the telescope diameter, such that $3~\lambda/D$ at 2.3~$\mu$m on a 10~m telescope corresponds to 14~AU for a star at 100~pc.} This puts the bulk of the giant planet population inside the \ed{IWA in the near-infrared, and limits the spectral coverage of direct imagers for planets close to their star. Interferometric techniques can nevertheless access smaller separations and therefore provide} the best opportunity for detecting \ed{giant} planets in the near-infrared. 

Vortex fiber nulling (VFN) is a single-aperture interferometric technique 
for detecting and characterizing faint companions at small separations 
\citep{Ruane2018_VFN,Echeverri2019_VFNOptLett,Ruane2019SPIE,Echeverri2021_BroadbandVFN}. \ed{VFN uses an optical vortex mask \citep{Beijersbergen1994} to impart a phase pattern that, when centered on a single-mode fiber, is orthogonal to the fiber's fundamental mode. Thus, a star can be aligned on-axis so that its light is rejected by the fiber while off-axis planet light couples in and is routed to a spectrograph for characterization.} VFN's simple optical design makes it easy to implement on existing and upcoming high-contrast imaging instruments with a fiber injection unit, thereby providing access to companions at ${\lesssim}1~\lambda/D$ (${\lesssim}5$~AU at 100~parsec for $\lambda{=}2.3~\mu$m and $D$=10~m). 
An on-sky VFN demonstrator is now operational~\citep{Echeverri2023_KPICVFNComm} as a new mode in the Keck Planet Imager and Characterizer ~(\citealp[KPIC -][]{Mawet2021,Delorme2021_KPIC,Echeverri2022_KPICPhaseII}; Jovanovic~et~al.~in~prep.) instrument at the Keck II Telescope. The nominal KPIC observing mode, referred to as direct spectroscopy (DS) since it aligns the fiber directly to the desired target, does not use a coronagraph and provides $R{\sim}35,000$ spectra that have been used extensively to spectroscopically characterize exoplanets and brown dwarf companions \citep{Wang2021_KPICScience, Wang2022_KPICCORetrieval, Wang2023_CORetrivalHR8799,Ruffio2023_KPICExomoons, Delorme2021_KPIC,Xuan2022_KPICRetrieval,Finnerty2023_WASP33B}. The new KPIC VFN mode builds on this to provide similar spectra for characterization at smaller separations. Additionally, since the VFN mode does not require prior knowledge of the exact position for the companion, it can be used to detect new companions. Previous commissioning results showed that, ignoring systematics such as fringing, the KPIC VFN mode's on-sky performance is sufficient for detecting companions 1000 times fainter than their host in K~band (2.0-2.4~$\mu$m) in 1 hour at separations of 30-80~mas
~\citep{Echeverri2023_KPICVFNComm}. 

In this paper, we now present the first detections from this new demonstrator mode. The three companions covered here were previously known only from RV and/or astrometric observations, such that our results represent their first direct detections and provide the first spectra for the companions.  
\ed{Though KPIC VFN alone can provide a detection,} in this paper we complement the VFN observations with CHARA observations using the MIRC-X and MYSTIC beam combiners~\citep{Anugu2020_MIRCXInstrument,Setterholm2023_MYSTICInstrument}, which have a demonstrated history of success at these angular separations~\citep[e.g.][]{Roettenbacher2015_CHARAOmiDra370:1,Roettenbacher2015_CHARASigGem270:1,thomas2021,DeFurio2022_MIRCXAStarSurvey,Lanthermann2023_MIRCXOStarSurvey}. \ed{This allows us to validate the VFN performance in this first demonstration against CHARA's well-established performance. For example, the CHARA data ensured} that the published orbital parameters are well-enough constrained that the targets were indeed within the current VFN field of view (${\sim}$30-80~mas) at the time of observation. \ed{Like this, if there were a VFN non-detection or anomalous result, we could be certain it was not due to the companion being too faint or beyond the VFN field of view. Moreover, the CHARA data provides complementary information to the VFN spectra. The latter} cannot constrain the companion position nor flux ratio, as the two parameters are degenerate \ed{in VFN's single annular coupling region. Thus, the CHARA results provide the first flux ratio measurements for the companions, highlighting some of the synergies between these long-baseline interferometry and vortex fiber nulling techniques}.

\section{Targets} \label{sec:Targets}
We targeted three nearby G stars with known companions at small separations. 
Table~\ref{tab:Observations} lists the targets and basic parameters of the primary star while the remainder of this section provides previously-known details on each target.

\movetabledown=2in
\begin{rotatetable}
\begin{deluxetable}{ccccccccccccc}
    \tablecaption{Targets and Observations\label{tab:Observations}}
    \tablehead{
        \colhead{Target} & \colhead{\ed{App.} Mag.} & \colhead{Spec.} & \colhead{Dist.} & \colhead{Pred. Flux} & \colhead{Date Observed} & \colhead{Instrument\ed{/Mode}} & \colhead{Obs.} & \colhead{Spectral} & \colhead{Int. Time} & \colhead{Pred. Sep.} & \colhead{Pred. RV} & \colhead{Bary. RV} \\[-6pt]
        \colhead{(HIP)} & \colhead{(K band)} & \colhead{Type} & \colhead{(pc)} & \colhead{Ratio (K)} & \colhead{(UT)} & \colhead{} & \colhead{Band} & \colhead{Resol.} & \colhead{(min)} & \colhead{(mas ; AU [; $\lambda/D$]\tablenotemark{\footnotesize{a}})} & \colhead{(km/s)} & \colhead{(km/s)}
    }
    \startdata
    21543 & 5.992 & G0 & 44.1 & 40-85 &
        2022 Oct 12 & \ed{KPIC/}VFN & K & 35,000 & 36\tablenotemark{\footnotesize{b}} & 48.7 ; 2.1 ; 1.0 & -7.9 & 22.8 \\[1mm]   
        & & & & & \multirow{2}{*}{2021 Oct 22} & \ed{CHARA/}MIRC-X & H & 50  & \multirow{2}{*}{10} & \multirow{2}{*}{20.5 ; 0.9 \phantom{; 1.0}} & & \\   
        & & & & &                              & \ed{CHARA/}MYSTIC & K & 50  & & & & \\[1mm]   
        & & & & & \multirow{2}{*}{2022 Sep 22} & \ed{CHARA/}MIRC-X & H & 50  & \multirow{2}{*}{28} & \multirow{2}{*}{47.3 ; 2.1 \phantom{; 1.0}} & & \\   
        & & & & &                              & \ed{CHARA/}MYSTIC & K & 100 & & & & \\[1mm]  
    \hline
    94666 & 6.280 & G0 & 64.4 & 105 &
        2023 May 09 & \ed{KPIC/}VFN & K & 35,000 & 48 & 37.3 ; 2.4 ; 0.8 & 13.5 & 14.1 \\[1mm]   
        & & & & & \multirow{2}{*}{2023 May 15} & \ed{CHARA/}MIRC-X & H & 50  & \multirow{2}{*}{9} & \multirow{2}{*}{38.1 ; 2.5 \phantom{; 1.0}} & & \\   
        & & & & &                              & \ed{CHARA/}MYSTIC & K & 100 & & & & \\[1mm]  
    \hline
    50319 & 4.345 & G0 & 31.2 & 405 &
        2023 May 06 & \ed{KPIC/}VFN & K & 35,000 & 123 & 48.7 ; 1.5 ; 1.0 & -9.2 & -28.3 \\[1mm]   
        & & & & & \multirow{2}{*}{2023 May 23} & \ed{CHARA/}MIRC-X & H & 50 & \multirow{2}{*}{46} & \multirow{2}{*}{41.9 ; 1.3 \phantom{; 1.0}} & & \\   
        & & & & &                              & \ed{CHARA/}MYSTIC & K & 50 & & & & \\[1mm]   
        & & & & & \multirow{2}{*}{2023 May 24} & \ed{CHARA/}MIRC-X & H & 50 & \multirow{2}{*}{37} & \multirow{2}{*}{41.5 ; 1.3 \phantom{; 1.0}} & & \\   
        & & & & &                              & \ed{CHARA/}MYSTIC & K & 50 & & & & \\   
    \enddata
    \tablenotetext{a}{$\lambda/D$ only provided for VFN observations, assuming $\lambda=2.3~\mu$m and $D=10$~m.}
    \tablenotetext{b}{Observations were made on two fibers, with 36 min per fiber. Only one fiber was used for the analysis.}
    \tablecomments{
    \ed{The first column is the Hipparcos number for the target. Second is the apparent K~band magnitude} from the 2MASS All-Sky Catalog. \ed{The spectral type is} from the Henry Draper Catalog and Extension. The distance, in parsecs, is derived from the parallax in the Gaia DR3 NSS table. The predicted \ed{K~band star-to-planet} flux ratio is computed and explained in Sec.~\ref{sec:Targets}. \ed{Columns six through nine provide the UT date of the observation, the instrument and observing mode, the astronomical band for the mode, and the corresponding spectral resolution ($R=\lambda/\Delta\lambda$).} Integration times \ed{in column ten} do not include calibrators, only the on-source time integrating for the companion. The predicted separation and RV at the time of observation use the Gaia NSS orbital solutions, except for HIP~21543 for which the \cite{Tokovinin2021_HIP21543} orbit was used. Pred. RV is the relative value between the primary and companion, and is only provided for the VFN data since CHARA cannot measure it. \ed{The final column is the average} Earth barycentric RV \ed{over the observation, and} is computed \ed{with the \texttt{Astropy} python package. This value is used to translate our measured RVs later in the paper from the instrument frame to the Earth-Sun barycenter so that values are reported with respect to that barycenter.}
    }
\end{deluxetable}
\end{rotatetable}

\textbf{HIP 21543} (HD~29310, vB~102) is a triple system in the Hyades cluster with an inner single-lined spectroscopic binary (SB1) \ed{first detected by \cite{Griffin1988_HIP21543OrigDetect}} and an outer visual companion \ed{originally detected at 0\farcs25 by \cite{Mason1993_HIP21543}}. The inner SB1 is the target of this paper. \cite{Tokovinin2021_HIP21543} combines RV observations with \ed{measurements of} the astrometric wobble of the outer companion (referred to as B) to provide a refined orbit for both the inner (referred to as Aa,Ab) and outer components. \ed{This puts Ab on a $734\pm0.3$ year orbit with a 37~mas semi-major axis, and B on a 125~year orbit with a 670~mas semi-major axis.} 
The mass ratio for Aa,Ab from their orbits is 0.29 such that given the estimated mass for Aa of $1.13~\Msun$, Ab is about $0.32~\Msun$. \cite{Bender2008_HIP21543SB2} reported weak lines from Ab, which would make this a double-lined spectroscopic binary with a direct detection, but 
\cite{Tokovinin2021_HIP21543} found that the measured RVs for the Ab lines are inconsistent with the astrometric wobble measurements. We note that the \cite{Tokovinin2021_HIP21543} orbits show Ab and B counter-orbiting around the central Aa star, implying an unusual orbital architecture for the system. 

An orbit for the inner Aa,Ab component is also reported 
in the Gaia DR3 non-single star (NSS) solutions \citep{gaiacollaboration_Gaia_2022, Holl2022_GaiaNSS}. The listed orbital period is $739\pm7$~days, consistent with the \cite{Tokovinin2021_HIP21543} value. From isochrone fitting, Gaia estimates the mass of Aa at $1.01\pm0.06~\Msun$ which allows them to predict the mass of Ab at $0.21\pm0.03~\Msun$ (see Gaia DR3 \texttt{binary\_masses} table, \citealt{gaiacollaboration_Gaia_2022}). With the Gaia-derived masses, we roughly estimate the $\Delta$K~mag between Aa and Ab. For Aa, we use the 2MASS K magnitude assuming it is dominated by the brighter primary star and neglecting its variability as a BY Draconis variable since the V band variability amplitude is only 0.03~mag \citep{lockwood_photometric_1984}, and likely even less in K band. Thus we estimate an absolute magnitude $M_\mathrm{K}=2.77$ given the Gaia parallax of 22.69~mas for the distance. For Ab, we use the latest version of the main-sequence dwarf table (MSDT) by \citet{pecaut_Intrinsic_2013} to estimate $M_\mathrm{K}\approx7.6$ assuming $0.21~\Msun$. This gives $\Delta\mathrm{K}\sim4.83$ for a flux ratio of ${\sim}85$ between the stars. A similar procedure but using the Tokovinin masses yields $\Delta\mathrm{K}\sim4.06$ (flux ratio ${\sim}40$).


\textbf{HIP 94666} (HD~180683) is also a triple system. There is an inner SB1 (Aa,Ab) with an orbital period of 1210~days provided by \cite{Tokovinin2018_MultiStarCatalog}. 
The outer visual companion, B, is on a ${\sim}$3000~year orbit at 3.6" \citep{Riddle2015_HIP94666,Roberts2017_HIP94666}. 
Though a full orbital solution is not provided in these prior works, the Gaia DR3 NSS table has a solution 
with a period of $1211\pm29$~days that is consistent with the published period for Aa,Ab. Gaia did not spectroscopically detect this system, so the \texttt{binary\_masses} table only provides a mass for Aa, $1.11_{-0.10}^{+0.06}~\Msun$. However, given the primary mass, we can solve for the mass ratio, $q$, between Ab/Aa using the Thiele-Innes orbital elements from \ed{the} Gaia DR3 NSS. We obtain $q\approx0.22$ which yields ${\sim}0.24~\Msun$ for Ab. 
We estimate the flux ratio as done for HIP~21543; 
the 2MASS K magnitude yields an absolute $M_\mathrm{K}=2.23$ for Aa given the 15.52~mas Gaia parallax and the MSDT yields $M_\mathrm{K}\approx7.3$ for Ab. 
We thus predict $\Delta\mathrm{K}\sim5.07$ (flux ratio ${\sim}105$).


\textbf{HIP 50319} (HD~89010, 35~Leo) is a SB1 binary (A,B) with an orbital period of ${\sim}$537~days~\citep{Tokovinin2014_HIP50319}. The Gaia NSS table again provides a full orbital solution, 
with a period of $524\pm6$ days. The \texttt{binary\_masses} table does not provide a mass for either star so we use the \cite{Tokovinin2014_HIP50319} mass of ${\sim}1.34~\Msun$ for A and the Gaia orbit to determine a mass ratio $q\approx0.11$, and hence a mass for B of ${\sim}0.15~\Msun$. As done for the other two targets, the 2MASS K magnitude gives an absolute $M_\mathrm{K}=1.88$ for A given the 32.09~mas Gaia parallax. From the MSDT, $M_\mathrm{K}\approx8.4$ for B, so we estimate $\Delta\mathrm{K}\sim6.52$ (flux ratio ${\sim}405$).


\section{Observations and Data Analysis} \label{sec:ObsAndAnalysis}
We observed all three targets with both KPIC VFN and CHARA MIRC-X/MYSTIC. VFN provides high-resolution spectra while CHARA gives the \ed{astrometry} and flux ratio. \ed{Note that the VFN spectra alone provide a detection, but for this first demonstration we complement the VFN results with CHARA observations to validate the VFN performance and highlight synergies with long-baseline interferometry. Below, we first summarize the observations from each instrument, and then describe the data reduction and analysis procedure.} 
Table~\ref{tab:Observations} lists observing parameters including dates, spectral band, spectral resolution, and integration time. It also lists the predicted flux ratio, separation, and relative RV at the time of observation based on the published orbits summarized above. 


\subsection{KPIC/VFN}
The VFN observations are done following a similar procedure to that presented in previous KPIC papers \citep[e.g.][]{Wang2021_KPICScience}. This involves: (1) observing a M giant to derive a wavelength solution, (2) observing a A0 standard star \ed{at a similar airmass as the target} to sample the telluric transmission, (3) observing the primary star, and then (4) observing the companion. 
However, in contrast to the direct spectroscopy (DS) observations where we offset the fiber to the companion in step four, in VFN \ed{mode the primary star is kept} on-axis but we insert a vortex mask so that it is nulled while the companion is preferentially coupled. \ed{Here we summarize the KPIC data reduction procedure using the KPIC DRP
; for details, see \citet{Wang2021_KPICScience}. First, we remove the thermal background and persistent bad pixels from the raw images by using instrument background frames taken before the observing night. Then, we use data from the telluric standard star to fit the trace of each column in the KPIC science fibers and nine spectral orders, which give the position and standard deviation of the PSF in the spatial direction at each column. For every frame, we extracted the 1D spectra in each column of each spectral order. To remove residual background light, we subtracted the median of pixels that are at least 5 pixels away from every pixel in each column. Finally, we used optimal extraction \citep{horne_optimal_1986} to sum the flux using weights defined by the 1D Gaussian line-spread function profiles calculated from spectra of the telluric star.} We only use KPIC echelle order 6 (${\sim}2.29$ to $2.34~\mu$m - \ed{correspondingly} NIRSPEC order 33) in this paper since it covers the CO bandhead where we expect many strong absorption lines from the M~dwarf companions. Furthermore, this echelle order is close to the central wavelength of 2.225~$\mu$m where the vortex provides the deepest nulls~\citep{Echeverri2023_KPICVFNComm}.  

\begin{figure}
    \begin{center}
    \includegraphics[width=\linewidth]{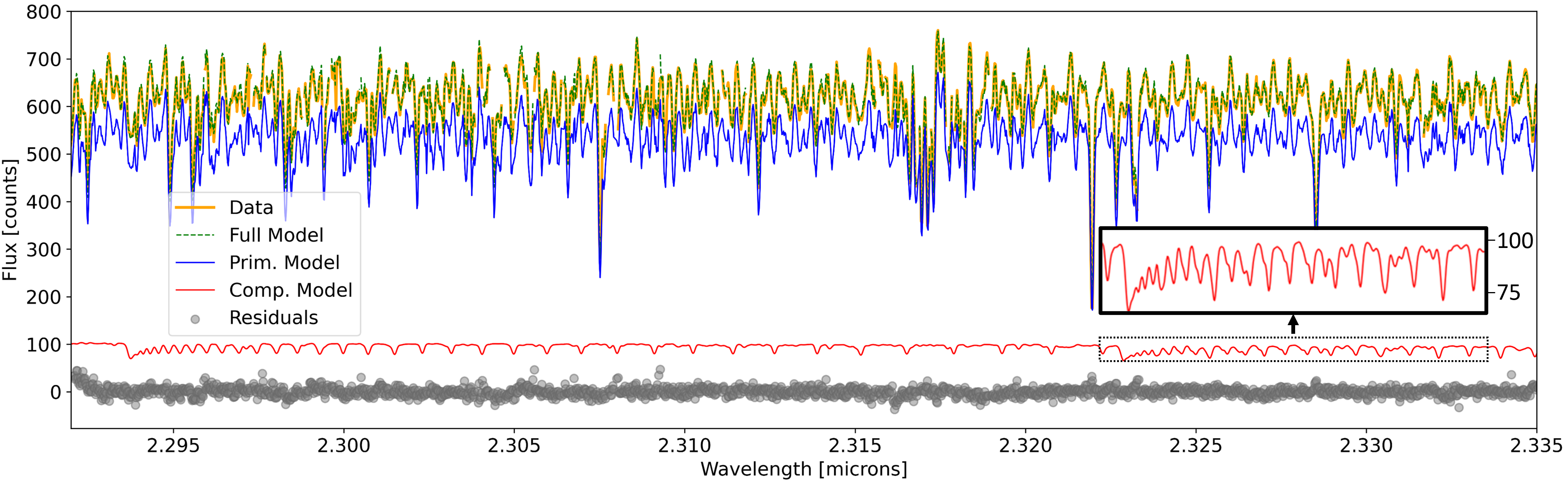}
    \end{center}
    \caption{
    Spectrum at R${\sim}35,000$ from VFN observations of HIP~21543. Only KPIC echelle order 6
    , the one used for the forward model fits, is shown. The raw spectrum has been median-filtered to remove the continuum and is \ed{shown as the} solid \ed{gold line,} while best-fit full model is in dashed \ed{green}. 
    Various components of the best-fit model are also shown: primary star spectrum (blue), companion spectrum (red), and residuals (grey dots). The companion spectrum is about five times fainter than that of the primary after partial nulling of the latter by VFN. The true flux ratio for the system is $70\pm11$, as measured by CHARA and shown in Table~\ref{tab:FitParametersCHARA}.
    \label{fig:HIP21543Spec}} 
\end{figure} 

The data analysis procedure \ed{for VFN is nearly-identical to that used for KPIC DS observations \citep[e.g.][]{Wang2021_KPICScience}. In short, we build a forward model of the data from a linear combination of the residual primary star flux and the companion flux. We account for the telluric and instrumental response using the A0 star spectrum.} To \ed{account for the residual light from} the primary star, we use the empirical spectra from the on-axis DS observations taken in step three above. This assumes the companion signal is negligible compared to the primary star signal since the companion is significantly fainter and less-efficiently coupled. For the companion model, we interpolate over a grid of BT-Settl (CIFIST) models \citep{Allard2012}, varying effective temperature (\Teff) and surface gravity (\logg) while assuming solar metallicity. Additionally, we fit for the RV shift and projected rotational rate (\vsini) of the companion. \ed{In KPIC data, a systematic fringing effect is introduced by Fabry-Perot cavities from transmissive optics in the instrument \citep{Finnerty2022_KPICPhaseI+Fringing}. To account for this fringing, we use the semi-physical fringing model described in \citet{Xuan2024} to model its effect in the data. This step is particularly important for VFN data since the characteristic fringing amplitude of ${\sim}1{-}3\%$ \citep{Xuan2024} caused by the residual primary starlight can be comparable to the companion signal in VFN observations.}
As a visual example of the elements that go into the forward model, Figure~\ref{fig:HIP21543Spec} shows the observed VFN mode spectrum for HIP~21543 along with the best-fit model and its various components. \ed{Furthermore, we carry out separate fits} to DS mode (ie. no nulling) spectra of the primary stars \ed{also using the BT-Settl models. The goal of these fits is to estimate the primary star's RV at the time of observation, and calculate the relative RV, which we} compare in Sec.~\ref{sec:results} to the expected relative RVs from the published orbits. \ed{Note that unlike for the VFN data, where we fit a combination of primary and companion flux, we only need to account for a single stellar component when fitting the primary star spectra.}

\ed{In addition to the spectral fits described above, we carry out a cross-correlation function (CCF) analysis to visualize the detection strength of the companion signal in our data (Fig.~\ref{fig:CCFs}). For the CCF analysis, we fix the companion template to the best-fit model from the spectral fit. Then, we estimate the maximum likelihood value for both the companion flux and speckle flux in the data as a function of RV shift, following \citet{Ruffio2019} and \citet{Wang2021_KPICScience}. The resulting estimate of the companion flux as a function of RV is the CCF. 
To estimate the signal-to-noise ratio (SNR) of the detection, we compare the the peak in the CCF to the standard deviation of the wings out to $\pm1000$~km/s; we report this value as the CCF SNR.}

\ed{Though the VFN data provides spectra which constitute a detection on their own and can be used for characterization, they do not provide reliable flux ratio measurements. The derived \Teff for the companion could be used to estimate the companion luminosity but this would be highly model-dependent. 
Robustly constraining flux ratios would require photometric flux measurements that are contingent on knowing the throughput to the detector. With VFN, the fiber coupling efficiency, and hence throughput, for the companion light depends on the angular separation to the center of the fiber. However, VFN's single annular fringe does not provide any spatial information for the companion. This does not limit our detection capabilities but does prevent us from determining the observed flux from VFN observations alone. For these targets, we instead use CHARA to constrain the position and flux ratios directly.}


\subsection{CHARA/MIRC-X and MYSTIC}
The Michigan InfraRed Combiner - eXeter \citep[MIRC-X -][]{Anugu2020_MIRCXInstrument} and the Michigan Young STar Imager \citep[MYSTIC -][]{Setterholm2023_MYSTICInstrument} on the Georgia State University Center for High Angular Resolution Astronomy (CHARA) Array \citep[][]{CHARA2005} were used to search for binary companions to all three targets. HIP~94666 and HIP~50319 were observed in 2023 specifically for a brief VFN follow-up program, with the latter target being observed over two nights. HIP~21543 had been observed twice in the past for other programs, so we used these archival data. \ed{The observations generally used all six one-meter telescopes in the array, with baselines spanning 30m to 330m, to provide an angular resolution down to $\frac{\lambda}{B_{\rm max}}\sim1$~mas, although only 5 telescopes were available for the HIP~89010 observations.  MIRC-X and MYSTIC were using their 6-beam All-in-One Combiners, providing up to 15 simultaneous baselines and 20 closure phases.}  Simultaneous MIRC-X and MYSTIC data were taken for each observation, with MIRC-X operating in H~band (1.50-1.72$\mu$m) and MYSTIC in K~band (1.95-2.38$\mu$m). \ed{Observing sequences involved interspersing target observations with calibrators to correct for the time-varying instrumental transfer function.} See Table~\ref{tab:Observations} for additional observing information. 

We reduce the interferometric data with the public {\texttt mircx\_pipeline} \citep{Anugu2020_MIRCXInstrument} \ed{to produce raw visibilities and closure phases.}  We then calibrate the transfer function using the calibrator stars, estimating their size using {\texttt Search Cal} \citep{chelli2016}. Then we look for a binary companion using a simple grid search, fitting only to the closure phases, while fixing the diameter of the primary estimated from photometry (0.26~mas for HIP~21543 and 0.53~mas for HIP~50319; irrelevant for HIP~94666 due to non-detection with CHARA). \ed{We note that closure phases are based on the sum of phases around closed triangles of baselines and are relatively free from calibration systematics that affect the visibility amplitudes \citep{monnier2007}.}  MIRC-X and MYSTIC each had different spectral resolutions and thus different interferometric fields-of-view and contrast sensitivities. Coupled with varying seeing conditions and different total observing times, there are some nights for which we are unable to recover reliable companion detections with both instruments. For the results reported below, we have applied the final wavelength correction terms found in \citet{torres2022}. With such a limited ``pilot program'' dataset, our error analysis is simplified, estimating position errors using the shape of the chi-squared surface immediately surrounding the best-fit companion position (see Fig.~\ref{fig:CCFs}), while upper limits on contrasts are derived from the contrast ratios from the best-fitting noise peaks. 

\section{Results and Discussion} \label{sec:results}
The VFN observations yielded confident detections on two of the companions while the third, HIP~50319~B, gave a tentative detection. Meanwhile, the CHARA observations yielded two confident detections and one non-detection, HIP~94666~Ab. \ed{Using the extracted KPIC spectra,} we make a first pass here at characterizing the companions to showcase the science capabilities of VFN, especially when combined with the input from CHARA. Tables~\ref{tab:FitParametersVFN} and~\ref{tab:FitParametersCHARA} summarize the best-fit values derived from the VFN and CHARA observations, respectively. 
The VFN fits fail to properly constrain \logg for the companions, which is partly due to the relatively low SNR and small wavelength coverage \ed{used for this VFN demonstration} (${\sim}2.29{-}2.34~\mu$m). In addition, constraining fundamental properties of M dwarfs, such as \logg and \Teff, is a challenging task and still remains somewhat model-dependent \citep[e.g.][]{Rajpurohit2018}. \ed{For example, the atmospheric models used to estimate \Teff and \logg may include inaccurate opacity data from outdated line lists, and insufficient treatment of dust opacity in late-type M dwarfs \citep{Iyer2023, Sanghi2023}. Furthermore, the BT-Settl grid we use has a coarse grid spacing of 0.5 dex in \logg, which could introduce interpolation issues \citep{Zhang2021}.} \ed{Despite these challenges, our derived \Teff for two of the targets was close to the expected value, and only the HIP~94666~Ab temperature seems to significantly deviate from expectation as explained below.} Thus, the VFN data provide the first spectra for the companions and the first constraints on their RV, \vsini, and \Teff. Given the high amplitude of residual primary flux in VFN spectra, the fringing signal is strong\ed{, and remains the dominant source of error in our spectral fits.} 
Beyond the VFN results, the CHARA results provide the first flux ratio and direct position measurements for the companions.  
\begin{deluxetable}{ccccccc}
    \tablecaption{Fitted Parameters from VFN\label{tab:FitParametersVFN}}
    \tablehead{
        \colhead{\makecell{Target \\ (HIP)}} &
        \colhead{\makecell{MJD}} &
        \colhead{\makecell{Prim. RV \\ (km/s)}} &
        \colhead{\makecell{Comp. RV \\ (km/s)}} &
        \colhead{\makecell{Rel. RV \\ (km/s)}} &
        \colhead{\makecell{\Teff \\ ($K$)}} &
        \colhead{\makecell{\vsini \\ (km/s)}} 
    }
    \startdata
    21543 & 59864.52 & $37.0^{+0.5}_{-0.6}$ & $45.3^{+0.4}_{-0.5}$ & $-8.3^{+0.6}_{-0.8}$ & $3480^{+90}_{-70}$ & $9.7^{+2.1}_{-0.9}$ \\
    94666 & 60073.62 & $-10.2^{+0.2}_{-0.2}$ & $-24.6^{+0.6}_{-0.5}$ & $14.4^{+0.6}_{-0.5}$ & $4090^{+320}_{-230}$ & $<7.2$\tablenotemark{\footnotesize a}  \\
    50319 & 60070.29 & $-35.4^{+0.1}_{-0.1}$ & $-17.2^{+0.9}_{-0.8}$ & $-18.2^{+0.9}_{-0.8}$ & $3300^{+130}_{-140}$ & $<10.1$\tablenotemark{\footnotesize a}  \\
    \enddata
    \tablenotetext{\footnotesize a}{Upper limit set at $2\sigma$.}
    \tablecomments{
    MJD is the average value during the observation. Primary and companion RV values are with respect to the Earth-Sun barycenter using the barycentric RV correction from Table~\ref{tab:Observations}. \Teff and \vsini in this table are for the companion.
    }
\end{deluxetable}
\begin{deluxetable}{cccccccccc}
    \tablecaption{Fitted Parameters from CHARA\label{tab:FitParametersCHARA}}
    \tablehead{
        \colhead{} & \colhead{} & \colhead{} & \colhead{} & \colhead{} & \colhead{} & \colhead{} & \multicolumn{3}{c}{Error Ellipse} 
        \\
        \cline{8-10}
        \colhead{\makecell{Target\\(HIP)}} & \colhead{MJD} & \colhead{Instrument} & \colhead{\makecell{Flux\\Ratio}} & \colhead{\makecell{Obs.\\Band}} & \colhead{\makecell{Sep.\\(mas)}} & \colhead{\makecell{PA\\(E of N)}} & \colhead{\makecell{Major Ax.\\(mas)}} & 
        \colhead{\makecell{Minor Ax.\\(mas)}} &
        \colhead{\makecell{PA of Major\\Ax. ($^\circ$)}}
    }
    \startdata
    21543 & 59509.323 & MIRC-X & 73 & H & 18.95 & 31.83 & 0.15 & 0.10 & 313 \\ 
     & & MYSTIC & 59 & K & 18.88 & 32.28 & 0.10 & 0.07 & 326 \\ 
     & 59844.485 & MIRC-X & 75 & H & 50.24 & 170.35 & 0.12 & 0.08 & 344 \\ 
     & & MYSTIC & 81 & K & 50.43 & 170.37 & 0.70 & 0.50 & 66 \\ 
    94666 & 60079.470 & MIRC-X & $>70$ & H & \nodata & \nodata & \nodata & \nodata & \nodata \\
     & & MYSTIC & $>40$ & K & \nodata & \nodata & \nodata & \nodata & \nodata \\
    50319 & 60087.201 & MIRC-X & $>180$ & H & \nodata & \nodata & \nodata & \nodata & \nodata \\ 
     & & MYSTIC & 407 & K & 56.83 & 71.64 & 0.26 & 0.11 & 69 \\ 
     & 60088.201 & MIRC-X & $>200$ & H & \nodata & \nodata & \nodata & \nodata & \nodata \\ 
     & & MYSTIC & 451 & K & 57.00 & 71.81 & 0.69 & 0.31 & 65 \\ 
    \enddata
\end{deluxetable}
\begin{figure}
    \begin{center}
    \includegraphics[width=\linewidth]{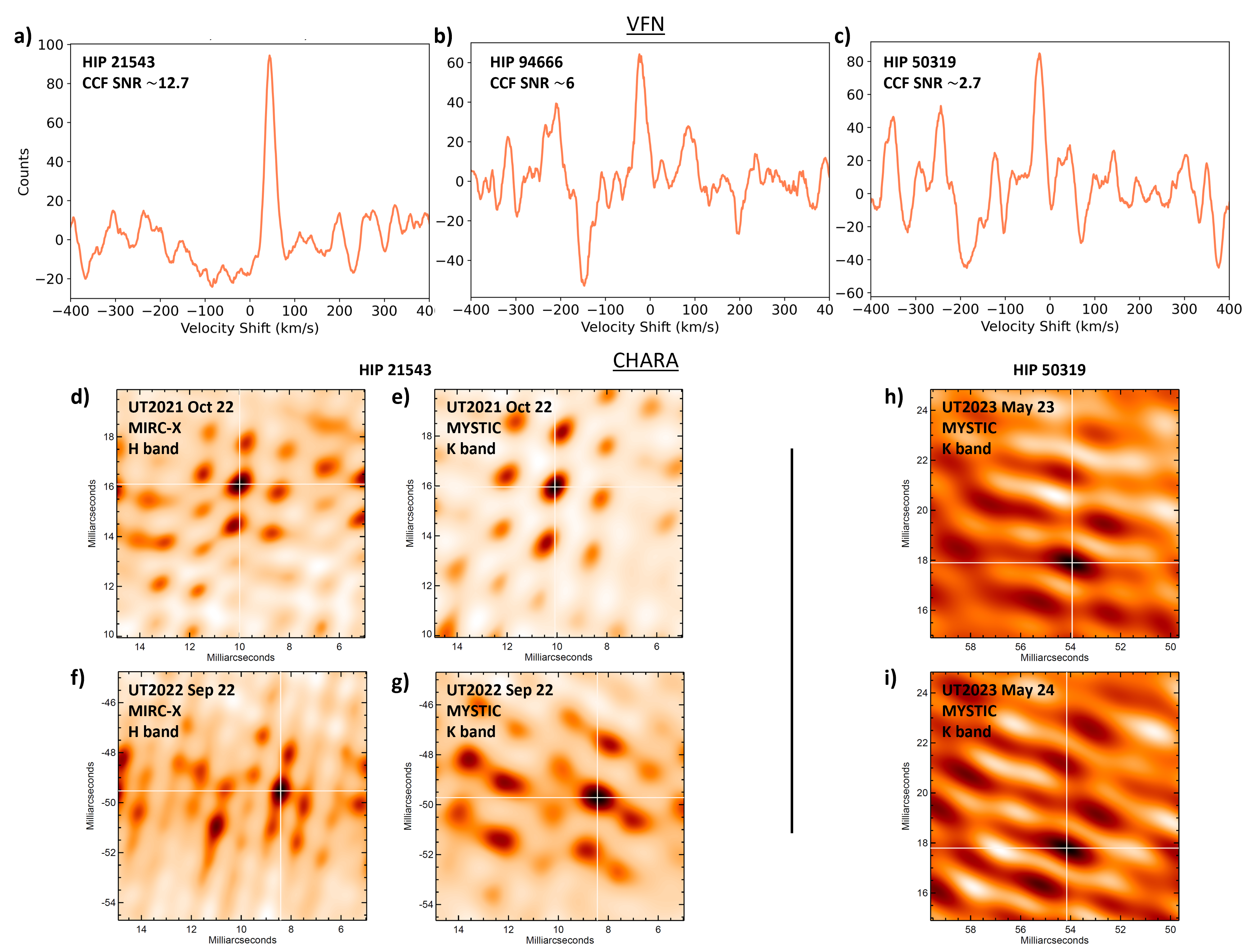}
    \end{center}
    \caption{
    Top row: CCF between the best-fit model and the measured spectra for (a) HIP~21543, (b) HIP~94666, and (c) HIP~50319. The CCF SNR is included in the top left of the plots. The periodic oscillations in the CCF for HIP~50319 are due to residual fringing that was not fully removed in the fits, limiting us to a tentative detection on this target. 
    Lower two rows: CHARA detection maps for HIP~21543 and HIP~50319. HIP~94666 is omitted as it was not detected by CHARA. The axes mark distance in milliarcseconds from the primary, with North up and East left. The white crosshairs denote the detected companion. Four maps are shown for HIP~21543, two for MIRC-X (d,f) and two for MYSTIC (e,g), across both nights. HIP~50319 has two maps (h,i), both from MYSTIC, one for each night. The upper left text in the CHARA maps denotes the observing night, beam combiner, and band for each plot. \label{fig:CCFs}} 
\end{figure} 

\textbf{HIP~21543} shows a strong detection with both instruments. \ed{Figure~\ref{fig:HIP21543Spec} shows the VFN mode spectrum, the best-fit full model, the on-axis DS mode spectrum used as the contribution from the primary in the model, and the resulting best-fit companion spectrum.} The CCF between the best-fit model and the VFN data is shown in Fig.~\ref{fig:CCFs}(a), yielding a CCF SNR of 12.7. 
The best-fit spectrum for the companion clearly shows two CO bandheads at around $2.295$ and $2.322~\mu$m. From this spectrum, we measure a companion RV of $37.0^{+0.5}_{-0.6}$~km/s. The fits to the DS mode spectra (not shown) give a primary RV of $45.3^{+0.4}_{-0.5}$~km/s. This yields a relative RV of $-8.3^{+0.6}_{-0.8}$~km/s between the primary and the companion on UT 2022 October 12, which agrees with the expected value of -7.9~km/s from the Tokovinin orbit and is close to the -6.9~km/s from the Gaia orbit. The best-fit \Teff just over $3450~K$ for Ab is slightly higher than, though still consistent with, the expected $3200-3300~K$ from the MSDT given the mass estimates. CHARA detected Ab with both MIRC-X and MYSTIC on both nights, as shown in Fig.~\ref{fig:CCFs}(d-g). The resulting separations of 18.9~mas and 50.3~mas for the two epochs are consistent with the expected values from both the Tokovinin orbit (20.5 and 47.3~mas) and the Gaia orbit (22.0 and 52.8~mas). 
The CHARA data also yield a K~band flux ratio from MYSTIC of $70\pm11$ between Ab and Aa, which is between the expected flux ratios of 42 and 85, again from the Tokovinin and Gaia masses respectively. 

\textbf{HIP~94666} has a confident detection in the KPIC VFN data with a CCF SNR of 6 as shown in Fig.~\ref{fig:CCFs}(b). \ed{Constraining the \Teff proved most challenging for this target out of the three}. \ed{Our fits give $4090^{+320}_{-230}~K$}, which is $800~K$ higher than the expected \ed{$\Teff\sim3250~K$ from the Gaia-derived companion mass of $\sim0.24~\Msun$. Part of the discrepancy could be due to underestimation in the Gaia mass ratio (and thereby flux ratio), for example if the secondary is bright enough to cause line blending \citep{Tokovinin2023_MultiSystems+NSS}. However, the CHARA non-detection puts a lower limit to the flux ratio of $>40$ in $K$ band, so line blending might be unlikely. On the other hand, our derived \Teff from the BT-Settl atmospheric models may be incorrect, as commonly seen in previous works on late M dwarfs \citep{Sanghi2023, Xuan2024}. For early M dwarfs like HIP~94666~Ab, however, \Teff accuracies of $100~K$ or lower have been achieved \citep[e.g.][]{Neves2014, Cristofari2022}, which may indicate that our fits are being biased by residual fringing from the bright primary star.} \ed{Despite the challenges with the \Teff,} our relative RV of $14.4^{+0.6}_{-0.5}$~km/s on 2023 May 9 is close to the predicted value of 13.5~km/s from the Gaia orbital solution. The CHARA observation yielded unreliable values in the short amount of integration time provided, such that we cannot provide position values and can only set lower limits on the flux ratio for the companion. We plan to re-observe this target with CHARA in 2024, \ed{which could inform on the true flux ratio and help resolve our discrepancy on the temperature}.

\textbf{HIP~50319} yielded a tentative VFN detection, with a CCF SNR of 2.7. The CCF, shown in Fig.~\ref{fig:CCFs}(c), has prominent structure in the wings, reflecting the fact that the detection is primarily limited by residual fringing. However, the best-fit model provides several pieces of evidence supporting the validity of this detection. First, the best-fit RV of ${-}35.4\pm0.1$~km/s for the primary on 2023 May 6 is in-line with the published velocity of ${\sim}$~-34~km/s~\citep{Deka-Szy2018_PTPS-RV,Nordstrom2004_GenevaCopenhagenRV}. Our fit to the primary further gives a $\Teff$ of $5480\pm10~K$ and $\vsini$ of $3.9\pm0.2$~km/s, which are close to the published values of $5686\pm7~K$ \citep{Deka-Szy2018_PTPS-RV} and $5.5$~km/s \citep{Luck2017_Abundances}.
Meanwhile, the fits to the companion spectrum show a RV of -$17.2^{+0.9}_{-0.8}$~km/s with a $\Teff\approx3300~K$. This $\Teff$ is close to the expected value of around $3000~K$ from the MSDT given the estimated mass. The fact that the retrieved properties for the primary are in-line with prior measurements, and that the companion RV and $\Teff$ are so different, provide strong evidence that our analysis \ed{of the VFN mode spectra} is indeed detecting spectral lines from two distinct objects.

The VFN-measured relative RV of $-18.2^{+0.9}_{-0.8}$~km/s is two times larger than expected from the Gaia orbital solution\ed{, which predicts a relative RV of $-9.2\pm3.0~$km/s at the time of the KPIC observations. It is possible that our VFN relative RVs for this system are biased by residual fringing in the data. Acquiring higher SNR spectra without fringing (see Sec.~\ref{sec:Conclustion}) could help confirm this. Alternatively, the Gaia orbit may not be entirely accurate.} The CHARA MYSTIC observations, shown in Fig.~\ref{fig:CCFs}(h,i), yielded confident detections that put the companion at a separation of around 56.9$\pm$0.3~mas for the two consecutive nights. \ed{This is $1\sigma$ higher than the Gaia-predicted separation of $42\pm14$~mas for the CHARA observation epoch, supporting a possible error in the Gaia orbital solution.} 
Similar discrepancies between ground-based RVs and Gaia NSS orbital solutions have been found in other studies \citep{Winn2022, Tokovinin2023_MultiSystems+NSS}, \ed{and have been attributed to incorrect orbital inclinations \citep{Marcussen2023}. However, the exact cause of the issue is unclear since only orbital solutions, and not the time-series astrometry, are published in Gaia DR3.} The CHARA MYSTIC detections provide a K~band flux ratio of $429\pm22$, which agrees with the predicted value of 405 presented in Sec.~\ref{sec:Targets}, \ed{implying that the Gaia mass ratio and companion mass of $0.11\Msun$ is likely accurate.} The MIRC-X data were unable to constrain the separation and only provided lower limits for the H~band flux ratio. 

Thus, the CHARA MYSTIC detection confirms that the companion was within the VFN field of view and should have been detectable at the time of observation. It also shows that the published orbital solution likely has errors that could explain the larger-than-expected relative RV from VFN. This, combined with the measured \Teff of the primary and companion, 
suggests a promising KPIC VFN detection of HIP~50319~B. 



\section{Conclusion} \label{sec:Conclustion}
In this paper we presented the first direct detections of three close-in low-mass stellar companions previously only known through indirect methods. 
\ed{The first two targets where confidently detected by VFN with CCF SNRs of 6 and 12.7.} Meanwhile, for the most-challenging target, 
our VFN detection is tentative due to strong fringing which could not be fully fitted and removed. An upgrade to KPIC in February 2024 will replace the optics that introduce fringing, significantly reducing the effect of this error in future observations. We will also add a new vortex mask, which will further improve the SNR by doubling the off-axis throughput and pushing the peak coupling from 1.4~$\lambda/D$ to 0.9~$\lambda/D$. Nevertheless, the current performance is sufficient for detection and characterization, as we are able to retrieve effective temperatures, rotational velocities, and RV values for the companions that are generally consistent with expectations. These VFN detections were made at separations between 35 and 55~mas, corresponding to around 0.7-1.2~$\lambda/D$ and about 2~AU. That is well within the typical IWA of conventional coronagraphs at these wavelengths, highlighting the power of cross-aperture nulling.

Previous single-telescope interferometric techniques have generally shown on-sky contrast limits of ${\sim}1500$ at ${\lesssim}2.5~\lambda/D$ \citep{Gauchet2016_NaCoNRMDetectLimits,Sallum2019_KeckNRM}, leading to demonstrated companion detections at flux ratios of a few-hundred within 2~$\lambda/D$ \citep{Hinkley2015_NRM7Detect,Lloyd2006_NRM80:1GJ802,Biller2012_SAM80:1}. A prior cross-aperture fiber nuller detected $\eta$~Peg~B with a flux ratio of 100 and measured the stellar diameter of the primary at a flux ratio of ${\sim}2000$ \citep{Serabyn2019_PFN}. 
Our previous VFN paper predicted contrast limits of ${\sim}1000$ at ${\sim}1~\lambda/D$ \citep{Echeverri2023_KPICVFNComm}, and this paper now adds companion detections with flux ratios around 100 and a tentative detection at ${\sim}430$. These VFN results also represent the first companion detection at these contrast levels with high ($R{>}10,000$) spectral-resolution nulling on-sky, showcasing the power of combining nulling interferometry with high-resolution spectroscopy and complementing the capabilities of previous instruments. In addition, these results are obtained at or within the conventional diffraction limit.

To \ed{validate our VFN detections with a well-established technique,} this paper combined KPIC VFN results with observations from CHARA. \ed{CHARA had confident detections on two of the targets, including the most challenging one that was tentative for VFN. }This allowed us to verify that we are close to our VFN contrast predictionsfrom \citet{Echeverri2023_KPICVFNComm}. Moreover, \ed{our combined results} highlight the complementary nature \ed{between long-baseline interferometry and cross-aperture nulling techniques. For example,} 
the CHARA-provided positions substantiate the published orbits, especially when combined with relative RV values from KPIC VFN. 
We find that the published orbits for the first two targets, HIP~21543~Ab and HIP~94666~Ab, are consistent with our results while the orbit for the third, HIP~50319~B, likely needs to be updated. 
These results open the door to detecting faint companions around young stars at separations within the IWA of typical coronagraphic imagers. \ed{Finally, they also point to an observing scheme that leverages the individual capabilities of VFN and CHARA.} Future surveys with CHARA and VFN can target young stars with Gaia-Hipparcos astrometric accelerations indicative of substellar companions. These interferometric surveys would complement imaging surveys already underway \citep[e.g.][]{Currie2021_AccelSurvey,Kuzuhara2022_AccelDetect,DeRosa2023_AccelDetect} to provide astrometric, flux ratio, and high-resolution spectral measurements at smaller separations, for a more efficient and complete view of faint, close-in companions. 

\section*{Acknowledgements} 
D.E was supported by a NASA Future Investigators in NASA Earth and Space Science and Technology (FINESST) fellowship under award \#80NSSC19K1423. D.E also acknowledges support from the Keck Visiting Scholars Program (KVSP) to install the KPIC Phase II upgrades required for KPIC VFN. J.X is supported by another FINESST award under \#80NSSC23K1434 and also acknowledges support from the KVSP to commission KPIC Phase II. 

Funding for KPIC has been provided by the California Institute of Technology, the Jet Propulsion Laboratory, the Heising-Simons Foundation (grants \#2015-129, \#2017-318, \#2019-1312, and \#2023-4598), the Simons Foundation (through the Caltech Center for Comparative Planetary Evolution), and the NSF under grant AST-1611623.

This work is based upon observations obtained with the Georgia State University Center for High Angular Resolution Astronomy Array at Mount Wilson Observatory.  The CHARA Array is supported by the National Science Foundation under Grant No. AST-1636624 and AST-2034336.  Institutional support has been provided from the GSU College of Arts and Sciences and the GSU Office of the Vice President for Research and Economic Development. S.K.\  and S.C.\ acknowledge funding for MIRC-X received from the European Research Council (ERC) under the European Union's Horizon 2020 research and innovation programme (Starting Grant No. 639889 and Consolidated Grant No. 101003096). J.D.M acknowledges funding for the development of MIRC-X (NASA-XRP NNX16AD43G, NSF-AST 1909165) and MYSTIC (NSF-ATI 1506540, NSF-AST 1909165).

The data presented herein were obtained at Keck Observatory, which is a private 501(c)3 non-profit organization operated as a scientific partnership among the California Institute of Technology, the University of California, and the National Aeronautics and Space Administration. The Observatory was made possible by the generous financial support of the W. M. Keck Foundation.

Some of this work was carried out at the Jet Propulsion Laboratory, California Institute of Technology, under a contract with the National Aeronautics and Space Administration and funded through the internal Research and Technology Development program.

This work has made use of data from the European Space Agency (ESA) mission
{\it Gaia} (\url{https://www.cosmos.esa.int/gaia}), processed by the {\it Gaia}
Data Processing and Analysis Consortium (DPAC,
\url{https://www.cosmos.esa.int/web/gaia/dpac/consortium}). Funding for the DPAC
has been provided by national institutions, in particular the institutions
participating in the {\it Gaia} Multilateral Agreement.

\vspace{3mm}
\noindent The authors wish to recognize and acknowledge the very significant cultural role and reverence that the summit of Maunakea has always had within the indigenous Hawaiian community. We are most fortunate to have the opportunity to conduct observations from this mountain.

%

\vspace{5mm}
\facilities{Keck:II (KPIC), CHARA (MIRC-X/MYSTIC)}


\software{KPIC DRP (\url{https://github.com/kpicteam/kpic_pipeline}), mircx\_pipeline (\url{https://gitlab.chara.gsu.edu/lebouquj/mircx_pipeline}), Search Cal (\url{https://www.jmmc.fr/english/tools/proposal-preparation/search-cal/}), Astropy \citep[\url{https://www.astropy.org/index.html} - ][]{astropy2022_V5.0,astropy2018_V2.0,astropy2013_V0.2}}






\bibliography{Libary}{}
\bibliographystyle{aasjournal}



\end{document}